\documentclass[twocolumn,showpacs,showkeys,amssymb]{revtex4}
\usepackage{graphicx}

\def\eV{\hbox{ eV}}
\def\GeV{\hbox{ GeV}}
\def\MeV{\hbox{ MeV}}
\def\Tr{\hbox{ Tr}}

\begin{document}

\title{Constraining Bilinear $R$-Parity Violation from Neutrino Masses}

\author{Marek G\'o\'zd\'z}
\email{mgozdz@kft.umcs.lublin.pl}
\author{Wies{\l}aw A. Kami\'nski}
\email{kaminski@neuron.umcs.lublin.pl}
\affiliation{
  Department of Informatics, Maria Curie-Sk{\l}odowska University \\
  pl. Marii Curie--Sk{\l}odowskiej 5, 20-031 Lublin, Poland}

\begin{abstract}
  We confront the $R$-parity violating MSSM model with the neutrino
  oscillation data. Investigating the 1--loop particle--sparticle
  diagrams with additional bilinear insertions on the external neutrino
  lines we construct the relevant contributions to the neutrino mass
  matrix. A comparison of the so-obtained matrices with the experimental
  ones assuming normal or inverted hierarchy and taking into account
  possible CP violating phases, allows to set constraints on the values
  of the bilinear coupling constants. A similar calculation is presented
  with the input from the Heidelberg--Moscow neutrinoless double beta
  decay experiment. We base our analysis on the renormalization group
  evolution of the MSSM parameters which are unified at the GUT scale.
  Using the obtained bounds we calculate the contributions to the
  Majorana neutrino transition magnetic moments.
\end{abstract}


\maketitle

\section{Supersymmetric model with $R$-parity violation}

The recent confirmation of neutrino oscillations \cite{nu-osc} gives
a~clear signal of existence of physics beyond the standard model of
particles and interactions (SM). Among many exotic proposals the
introduction of supersymmetry (SUSY) proved to be both elegant and
effective in solving some of the drawbacks of the SM. The minimal
supersymmetric standard model (MSSM) (a~comprehensive review can be
found in \cite{mssm}) populates the so-called desert between the
electroweak and the Planck scale with new heavy SUSY particles, thus
removing the scale problem. What is more, using the MSSM renormalization
group equations for gauge couplings indicates that there is
a~unification of $g_1$, $g_2$ and $g_3$ around $m_{GUT}\approx 1.2
\times 10^{16}\GeV$ which means that MSSM in a~somehow natural way
includes Grand Unified Theories (GUTs). This model is also characterized
by a~heavier Higgs boson, comparing with the Higgs boson predicted by
SM, which is in better agreement with the known experimental data. New
interactions present in MSSM lead to many exotic processes which opens
a~completely new field of research.

Building the minimal supersymmetric version of the Standard Model one
usually assumes the conservation of the $R$-parity, defined as
$R=(-1)^{3B+L+2S}$, where $B$ is the baryon number, $L$ the lepton
number, and $S$ the spin of the particle. The definition implies that
all ordinary SM particles have $R=+1$ and all their superpartners have
$R=-1$. In theories preserving $R$-parity the product of $R$ of all the
interacting particles in a~vertex of a~Feynman diagram must be equal to
$1$. This implies that the lepton and baryon numbers are conserved, and
that SUSY particles are not allowed to decay to non-SUSY ones. It
follows that the lightest SUSY particle (usually the lightest neutralino
$\tilde\chi^0_1$) must remain stable, giving a~good natural candidate
for cold dark matter. All this makes the $R$-parity conserving models
very popular.

In practice, however, the $R$-parity conservation is achieved by
neglecting certain theoretically allowed terms in the superpotential.
Casting such hand-waving approach away, one should retain these terms,
finishing with an~$R$-parity violating (RpV) model, with richer
phenomenology and many even more exotic interactions
\cite{aul83,valle,np,rbreaking}. The RpV models provide mechanisms of
generating Majorana neutrino masses and magnetic moments, describe
neutrino decays, SUSY particles decays, exotic nuclear processes like
the neutrinoless double beta decay, and many more. Being theoretically
allowed, RpV SUSY theories are interesting tools for studying the
physics beyond the Standard Model. The many never-observed processes
allow also to find severe constraints on the non-standard parameters of
these models, giving an insight into physics beyond the SM.

The violation of the $R$-parity may be introduced in a~few different
ways. In the first one $R$-parity violation is introduced as a
spontaneous process triggered by a non-zero vacuum expectation value of
some scalar field \cite{aul83}. Another possibilities include the
introduction of additional bi- \cite{valle} or trilinear
\cite{rbreaking} RpV terms in the superpotential, or both. In the
following we incorporate the explicit RpV breaking scenario.

The $R$-parity conserving part of the superpotential of MSSM is usually
written as
\begin{eqnarray}
  W^{MSSM} &=& \epsilon_{ab} [
  (\mathbf{Y}_E)_{ij} L_i^a H_u^b \bar E_j
  + (\mathbf{Y}_D)_{ij} Q_{ix}^{a} H_d^b \bar D_{j}^{x} \nonumber \\
  &+& (\mathbf{Y}_U)_{ij} Q_{i x}^{a} H_u^b \bar U_{j}^{x} + \mu H_d^a H_u^b], 
\label{wmssm}
\end{eqnarray}
while its RpV part reads
\begin{eqnarray}
  W^{RpV} &=& \epsilon_{ab}\left[
    \frac{1}{2} \lambda_{ijk} L_i^a L_j^b \bar E_k
    + \lambda'_{ijk} L_i^a Q_{jx}^{b} \bar D_{k}^{x} \right] \nonumber \\
  &+& \frac{1}{2}\epsilon_{xyz} \lambda''_{ijk}\bar U_i^x\bar
  D_j^y \bar D_k^z + \epsilon_{ab}\kappa^i L_i^a H_u^b.
\label{WRPV}
\end{eqnarray}
The {\bf Y}'s are 3$\times$3 Yukawa matrices. $L$ and $Q$ are the
$SU(2)$ left-handed doublets while $\bar E$, $\bar U$ and $\bar D$
denote the right-handed lepton, up-quark and down-quark $SU(2)$
singlets, respectively. $H_d$ and $H_u$ mean two Higgs doublets. We have
introduced color indices $x,y,z = 1,2,3$, generation indices
$i,j,k=1,2,3=e,\mu,\tau$ and the SU(2) spinor indices $a,b = 1,2$.

As far as the (in principle unknown) RpV coupling constants are
concerned, the most popular approach is to neglect the bilinear terms
and to discuss the effects connected with the trilinear terms only. In
such a~case, if one is not intereseted in exotic baryon number violating
processes, one has to additionally set $\lambda''=0$, which ensures the
stability of the proton. In this paper we concentrate on the bilinear
terms only and set all trilinear RpV couplings to zero.

For completeness we write down the scalar mass term present in our model,
\begin{eqnarray}
{\cal L}^{mass} &=& \mathbf{m}^2_{H_d} h_d^\dagger h_d +
                    \mathbf{m}^2_{H_u} h_u^\dagger h_u +
     q^\dagger \mathbf {m}^2_Q q + l^\dagger \mathbf {m}^2_L l \nonumber \\
&+&  u \mathbf {m}^2_U u^\dagger + d \mathbf {m}^2_D d^\dagger +
     e \mathbf {m}^2_E e^\dagger,
\end{eqnarray}
the soft gauginos mass term ($\alpha=1,...,8$ for gluinos)
\begin{equation}
  {\cal L}^{gaug.} = \frac12 \left( 
  M_1 \tilde{B}^\dagger \tilde{B} + 
  M_2 \tilde{W_i}^\dagger \tilde{W^i} +
  M_3 \tilde{g_\alpha}^\dagger \tilde{g^\alpha} + h.c.\right ),
\end{equation}
as well as the supergravity mechanism of supersymmetry breaking, by
introducing the Lagrangian
\begin{eqnarray}
  {\cal L}^{soft} &=& \epsilon_{ab} \Big[
  (\mathbf{A}_E)_{ij} l_i^a h_d^b \bar e_j
  + (\mathbf{A}_D)_{ij} q_i^{ax} h_d^b \bar d_{jx} \nonumber \\
  &+& (\mathbf{A}_U)_{ij} q_i^{ax} h_u^b \bar u_{jx} + B \mu h_d^a h_u^b +
  B_i \kappa_i l_i^a h_u^b \Big],
\end{eqnarray}
where lowercase letters stand for scalar components of the respective
chiral superfields, and 3$\times$3 matrices {\bf A} as well as $B\mu$
and $B_i$ are the soft breaking coupling constants.


\section{Neutrino--neutralino mixing}

The inclusion of the bilinear RpV terms imply mixing between neutrinos
and neutralinos. In the basis $(\nu_e,\nu_\mu,\nu_\tau,\tilde B,\tilde
W^3, \tilde H_d^0, \tilde H_u^0)$ the full $7\times 7$
neutrino--neutralino mixing matrix may be written \cite{np} in the
following form:
\begin{equation}
M_{\nu\tilde\chi^0} = 
\pmatrix{
  0_{3\times 3} & m             \cr 
  m^T           & M_{\tilde\chi^0}
},
\label{eq:nu-neutr-matrix}
\end{equation}
where 
\begin{equation}
m = 
\pmatrix{
  -\frac12 g' \omega_e    & \frac12 g \omega_e    & 0 & -\kappa_e   \cr
  -\frac12 g' \omega_\mu  & \frac12 g \omega_\mu  & 0 & -\kappa_\mu \cr
  -\frac12 g' \omega_\tau & \frac12 g \omega_\tau & 0 & -\kappa_\tau 
}
\end{equation}
and $M_{\tilde\chi^0}$ is the standard MSSM neutralino mass matrix:
\begin{equation}
M_{\tilde\chi^0} = 
\pmatrix{
  M_1 & 0   & -\frac12 g' v_1 &  \frac12 g' v_2 \cr
  0   & M_2 &  \frac12 g  v_1 & -\frac12 g  v_2 \cr
  -\frac12 g' v_1 &  \frac12 g v_1 & 0    & -\mu \cr
   \frac12 g' v_2 & -\frac12 g v_2 & -\mu &   0  }.
\end{equation}
The matrix (\ref{eq:nu-neutr-matrix}) has the seesaw--like structure and
contains the sneutrino vacuum expectation values (vevs) $\omega_i$. These are
in general free parameters which contribute to the gauge boson masses
via the relation
\begin{equation}
  v_1^2 + v_2^2 + \sum_{i=e,\mu,\tau} \omega_i^2 = v^2 =
  \left( \frac{2 M_W}{g} \right)^2 \simeq (246 \GeV)^2,
\end{equation}
where $v_1$ and $v_2$ are the usual down-type and up-type Higgs boson
vevs, respectively. By introducing the angle $\beta$ defined by
$\tan\beta=v_2/v_1$ we obtain four free parameters of the theory:
$\tan\beta$ and $\omega_i$.  Fortunately it turns out that in order to
obtain proper electroweak symmetry breaking the sneutrino vevs cannot be
arbitrary. We give the details in the next section.


\begin{figure*}
  \centering
  \includegraphics[height=0.8\textwidth,angle=270]{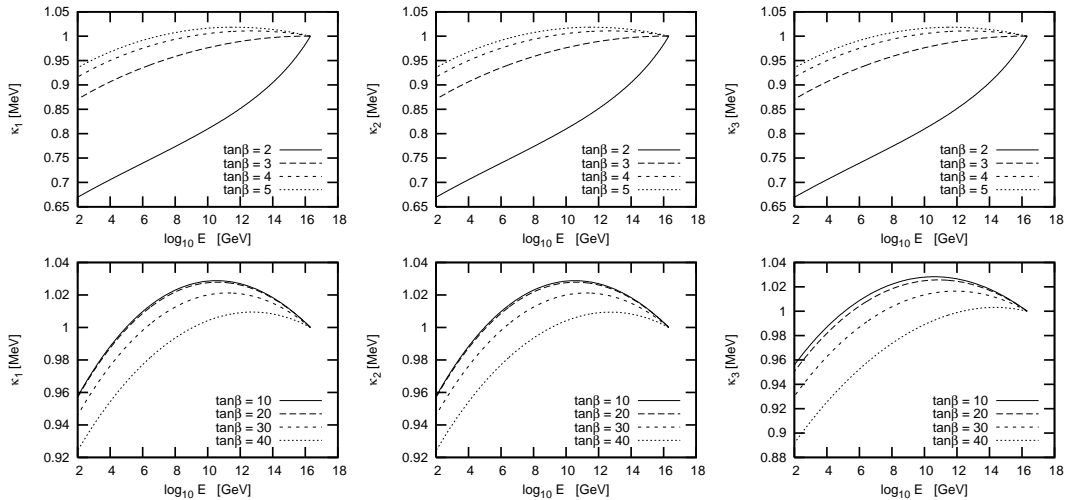}
  \caption{An example of RG running of the bilinear $\kappa_i$
    couplings. The unification scenario was: $m_0=200\GeV$,
    $m_{1/2}=500\GeV$, $A_0=200$, sgn$(\mu)=1$. All $\kappa_i^{GUT}$
    were equal to 1 MeV.}
  \label{fig:kappa}
\end{figure*}

\section{Handling the free parameters}

The RpV MSSM model introduces several new free parameters when compared
with the usual MSSM. Fortunately their number can be constrained by
imposing GUT unification and renormalization group evolution. In this
paper we restrict ourselves to the bilinear RpV couplings only, setting
all trilinear couplings ($\lambda$, $\lambda'$, $\lambda''$) to zero.
This assumption simplifies some of the RGE equations, which we list
below. Such approach leads at the end to the following set of free
parameters: $m_0$, $m_{1/2}$, $A_0$, $\tan\beta$, sgn$(\mu)$, and
$\kappa_i^{GUT}$ ($i=1,2,3)$.

\subsection{Masses and soft breaking couplings}
The masses of all the supersymmetric scalars are unified at $m_{GUT}$ to
a~common value $m_0$, and of all the supersymmetric fermions to
$m_{1/2}$. The values of the trilinear soft SUSY breaking couplings are 
set according to the following relations \cite{Hirsch}
\begin{eqnarray}
  \mathbf{A}_{E,D,U} &=& A_0 \mathbf{Y}_{E,D,U}, \\
  B = B_{1,2,3} &=& A_0 -1.
\end{eqnarray}
The RGE equations for the $\mathbf{A}$ couplings can be found elsewhere
\cite{kanekolda,drtjones,MartinVaughn,mg-art1,ChoMisiak,rge,rsugra}.
The $B$ couplings are evolved down to the low energy regime according to
the renormalization group equations
\begin{eqnarray}
  16\pi^2 \frac{dB}{dt} &=&
  6\Tr(\mathbf{A}_U\mathbf{Y}_U^\dagger) +
  6\Tr(\mathbf{A}_D\mathbf{Y}_D^\dagger)  \nonumber \\
  &+&
  2\Tr(\mathbf{A}_E\mathbf{Y}_E^\dagger) + 6 g_2^2 M_2 + 2 g_1^2 M_1, \\
  16\pi^2 \frac{dB_{1,2}}{dt} &=&
  6\Tr(\mathbf{A}_U\mathbf{Y}_U^\dagger) + 6 g_2^2 M_2 + 2 g_1^2 M_1, \\
  16\pi^2 \frac{dB_3}{dt} &=&
  6\Tr(\mathbf{A}_U\mathbf{Y}_U^\dagger)  \nonumber \\
  &+&
  2\Tr(\mathbf{A}_E\mathbf{Y}_E^\dagger) + 6 g_2^2 M_2 + 2 g_1^2 M_1.
\end{eqnarray}
where $g_1^2 = 5/3\ g'^2/(4\pi^2)$ and $g_2 = g^2/(4\pi^2)$, $5/3$ being
the GUT normalization factor.

\subsection{Bilinear $\kappa_i$ couplings}
The three $\kappa_i^{GUT}$ couplings at GUT scale remain free in our
model. After setting them the couplings are evolved down to the $m_Z$
scale according to the renormalization group equations which in our case
take the following simple form:
\begin{eqnarray}
  16\pi^2 \frac{d\kappa_i}{dt} 
  &=& \kappa_i (3\Tr(\mathbf{Y}_U \mathbf{Y}_U^\dagger) - 3 g_2^2 - g_1^2) 
  \nonumber \\
  &+& \sum_{j=1}^3 \kappa_j (\mathbf{Y}_E \mathbf{Y}_E^\dagger)_{ij}.
\end{eqnarray}

An example of the running of $\kappa_i$ is presented on
Fig.~\ref{fig:kappa}. One sees that for higher $\tan\beta$ the couplings
vary rather weakly (notice the logarithmic scale on the energy axis) for
the whole energy range between the electroweak scale $m_Z$ and
$m_{GUT}$. For small $\tan\beta<10$ the difference between the $m_{GUT}$
and $m_Z$ values are of the order of $\le 35\%$. The value 1~MeV at the
GUT scale was chosen arbitrarily; we will show later that this is the
typical order of magnitude for which agreement with experimental data on
neutrino masses and mixing may be obtained.

\subsection{Vacuum expectation values}
At the beginning of the numerical procedure we set the down and up Higgs
vevs to
\begin{equation}
  v_1 = v \cos\beta, \qquad v_2 = v \sin\beta,
\end{equation}
while the initial guess for the sneutrinos vevs is 
\begin{equation}
  \omega_i~=0.
  \label{eq:vevini}
\end{equation}
The actual values of $\omega_i$ are calculated from the condition that
at the electroweak symmetry breaking scale the linear potential is
minimized. By taking partial derivatives of the potential one obtains
the so-called tadpole equations \cite{Hirsch}, which are zero at the
minimum.


In our procedure we solve three equations, which can be written as
($i=1,2,3$)
\begin{equation}
  \kappa_i(v_1\mu - v_2B_i) = \sum_{j=1}^3 \omega_j~\Omega_{ji},
\end{equation}
where
\begin{equation}
  \Omega_{ji}=\kappa_j\kappa_i + (\mathbf{m}_L^2)_{ji} + \delta_{ji} D,
\end{equation}
$\delta_{ji}$ being the Kronecker delta, and
\begin{equation}
  D = \frac18 (g^2+g'^2) (v^2 - 2v_2^2).
\end{equation}
Notice that they are linear in $\omega_{1,2,3}$ and therefore this set
has only one solution. After finding it, we use the trigonometric
parameterization \cite{Hirsch}, which preserves the definition of $\tan\beta$,
\begin{eqnarray}
  v_1      &=& v \sin\alpha_1 \sin\alpha_2 \sin\alpha_3 \cos\beta, \\
  v_2      &=& v \sin\alpha_1 \sin\alpha_2 \sin\alpha_3 \sin\beta, \\
  \omega_1 &=& v \cos\alpha_1 \sin\alpha_2 \sin\alpha_3, \\
  \omega_2 &=& v \cos\alpha_2 \sin\alpha_3, \\
  \omega_3 &=& v \cos\alpha_3,
\end{eqnarray}
to calculate new values of $v_1$ and $v_2$. We return back to the
tadpoles with these new values and continue in this way until
self-consistency of the results is reached. It turns out that due to
the expected smallness of the $\omega_i$ vevs, the initial guess
Eq.~(\ref{eq:vevini}) is quite a~good approximation. It usually suffices
to repeat the whole procedure three times to obtain self-consistency at the
level of ${\cal O}(10^{-4})$, which is more than enough for our purposes.
The so-obtained set of vevs is used during the determination of the
mass spectrum of the model.


\section{Feynman diagrams with RpV couplings on the external neutrino
  lines}

It is well known that, once allowing for $R$-parity violation,
a~particle--sparticle 1--loop diagrams give important corrections to the
usual tree level neutrino mass term. These processes have been
extensively discussed in the literature
\cite{Haug,Bhatta,Abada,rpvneutrinos,mg-art6,mg-art9,mg-art11}, mainly
in the context of constraining the tree--level alignment parameters
$\Lambda$ or the trilinear RpV couplings $\lambda$ and $\lambda'$
\cite{bounds-3linear,mg-art6}.

In general, the explicit RpV effects may be taken into account in three
different ways. One may include the bilinear RpV couplings or the
trilinear couplings, or both. Of course the most complete one is the
third possibility, which is at the same time the most complicated.
Therefore it is customary to limit the discussion to either tri- or
bilinear terms only. In this paper we are interested in bilinear
couplings and set all trilinear couplings to zero.

In order to discuss the possible magnitude of the bilinear RpV couplings
$\kappa_i$ we extend the simplest diagrams by including the
neutrino--neutralino mixing on the external lines.

\begin{figure}[t]
  \centering
  \includegraphics[width=0.45\textwidth]{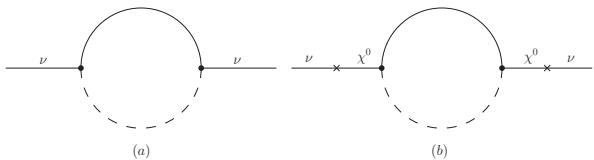}
  \caption{\label{fig:loop} (a) The basic 1--loop diagram giving
    rise to the Majorana neutrino mass in the $R$-parity violating MSSM. (b)
    1--loop diagram with RpV neutrino--neutralino couplings included on
    the external lines.}
\end{figure}

The topology of the basic type of 1--loop diagrams we will consider is
presented on Fig.~\ref{fig:loop}(a). These diagrams lead to Majorana
neutrino mass term, where the effective interaction vertex is expanded
into the RpV particle--sparticle loop. These diagrams and their more
complicated versions with the Higgs bosons and sneutrinos inside the
loop, were classified in e.g. Ref.~\cite{rpvneutrinos} and discussed in
details elsewhere (see
\cite{Haug,Bhatta,Abada,rpvneutrinos,mg-art6,mg-art9,mg-art11,bounds-3linear}
among others). In the present paper we add the possible
neutrino--neutralino mixing on the external lines
(Fig.~\ref{fig:loop}(b)), which leads to another contributions to the
neutrino mass.  Obviously this additional contribution must be in
agreement with the present experimental data. We discuss two main cases,
in which either lepton and slepton or quark and squark are in the loop
(in the case of higgsino $\tilde H_u$ only the up-type quarks count). At
the same time the neutrino may mix either with the gauginos: bino
$\tilde B$ or wino $\tilde W^3$, or with the neutral up-type higgsino
$\tilde H_u$. All the nine cases together with the relevant bi- and
trilinear coupling constants have been gathered in
Tab.~\ref{tab:1}.

\begin{figure}
  \centering
  \includegraphics[width=0.35\textwidth]{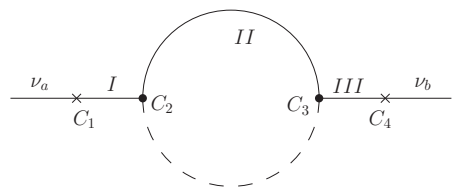}
  \caption{\label{fig:diag} Diagrams with bilinear neutrino--neutralino
    interactions leading to the Majorana neutrino mass.}
\end{figure}

\begin{table}
  \centering
  \caption{\label{tab:1} Nine diagrams with neutrino--neutralino mixing
    on the external lines leading to Majorana neutrino mass. $I$ and
    $III$ are the neutralinos which mix with the neutrinos on the external
    lines of the diagram depicted on Fig.~\ref{fig:diag}. $II$ defines
    the content of the loop ($u\tilde u$ stands for the up-type
    (s)quarks, $d\tilde d$ stands for the down-type (s)quarks,$q\tilde
    q$ for all (s)quarks, and $l\tilde l$ for (s)leptons).}
  \begin{tabular}{lccccccc}
    \hline\hline
    & $I$ & $II$ & $III$ & $C_1$ & $C_2$ & $C_3$ & $C_4$ \\
    \hline
    1 & $\tilde H_u$ & $u\tilde u$ & $\tilde H_u$ & 
    $\kappa_a$ & $\sqrt{2} m_u/v_u$ & $\sqrt{2} m_u/v_u$ & $\kappa_b$ \\
    2 & $\tilde H_u$ & $u\tilde u$ & $\tilde B$ &
    $\kappa_a$ & $\sqrt{2} m_u/v_u$ & $-g'/(3\sqrt{2})$ & $g'\omega_b$ \\
    3 & $\tilde H_u$ & $u\tilde u$ & $\tilde W^3$ &
    $\kappa_a$ & $\sqrt{2} m_u/v_u$ & $-g/\sqrt{2}$ & $g\omega_b$ \\
    4 & $\tilde B$   & $q\tilde q$ & $\tilde B$ &
    $g'\omega_a$ & $-g'/(3\sqrt{2})$ & $-g'/(3\sqrt{2})$ & $g'\omega_b$ \\
    5 & $\tilde B$   & $l\tilde l$ & $\tilde B$ &
    $g'\omega_a$ & $-g'/\sqrt{2}$ & $-g'/\sqrt{2}$ & $g'\omega_b$ \\
    6a & $\tilde W^3$ & $u\tilde u$ & $\tilde W^3$ &
    $g\omega_a$ & $-g/\sqrt{2}$ & $-g/\sqrt{2}$ & $g\omega_b$ \\
    6b & $\tilde W^3$ & $d\tilde d$ & $\tilde W^3$ & 
    $g\omega_a$ & $g/\sqrt{2}$ & $g/\sqrt{2}$ & $g\omega_b$ \\
    7 & $\tilde W^3$ & $l\tilde l$ & $\tilde W^3$ &
    $g\omega_a$ & $g/\sqrt{2}$ & $g/\sqrt{2}$ & $g\omega_b$ \\
    8a & $\tilde B$   & $u\tilde u$ & $\tilde W^3$ &
    $g'\omega_a$ & $-g'/(3\sqrt{2})$ & $-g/\sqrt{2}$ & $g\omega_b$ \\
    8b & $\tilde B$   & $d\tilde d$ & $\tilde W^3$ &
    $g'\omega_a$ & $-g'/(3\sqrt{2})$ & $g/\sqrt{2}$ & $g\omega_b$ \\
    9 & $\tilde B$   & $l\tilde l$ & $\tilde W^3$ &
    $g'\omega_a$ & $g'/\sqrt{2}$ & $g/\sqrt{2}$ & $g\omega_b$ \\
    \hline\hline
  \end{tabular}
\end{table}

The contributions from individual diagrams have been calculated using
the same technique as in Refs.~\cite{mg-art6,mg-art9}. In
Ref.~\cite{mg-art9} we have discussed the possible influence of
including the quark mixing in the calculations. Here we neglect this
effect. 

The neutrino mass matrix resulting from the bilinear processes only can
be written as the following sum:
\begin{equation}
  {\cal M}_{ab} = \sum_{i=1}^9 {\cal M}_{ab}^i,
\label{eq:M-full}
\end{equation}
where the separate contributions read
\begin{equation}
  {\cal M}_{ab}^i = \frac{1}{16\pi^2} 
  \frac{C_1 C_2 C_3 C_4}{m_I m_{III}} F_{II}.
\end{equation}
The masses of the neutralinos $m_I$ and $m_{III}$, and the coupling
constants have to be taken from Tab.~\ref{tab:1}. The functions $F$
represent the contributions from the particle--sparticle loops. They
read:
\begin{eqnarray}
  && F_{u\tilde u} =  \sum_{i,j}
  \left[3 \frac{\sin2\theta^j}{2} m_{u^i} f(x_2^{ij}, x_1^{ij}) \right],
\label{eq:Fu}\\
  && F_{d\tilde d} =  \sum_{i,j}
  \left[3 \frac{\sin2\theta^j}{2} m_{d^i} f(x_2^{ij}, x_1^{ij}) \right],
\label{eq:Fd}\\
  && F_{q\tilde q} = \sum_{i,j}
  \left[3 \frac{\sin2\theta^j}{2} m_{q^i} f(x_2^{ij}, x_1^{ij}) \right], 
\label{eq:Fq}\\
  && F_{l\tilde l} =  \sum_{i,j}
  \left[  \frac{\sin2\phi^j}{2} m_{l^i} f(y_2^{ij}, y_1^{ij}) \right],
\label{eq:Fl}
\end{eqnarray}
where $\theta^j$ and $\phi^j$ are the $j$-th squark and slepton mass
eigenstates' mixing angles, respectively. For simplicity we have defined
dimensionless quantities $x_{1,2}^{ab}=(m_{q^a}/m_{\tilde
  q_{1,2}^b})^2$, which are the $a$-th quark mass over the $b$-th squark
first or second mass eigenstate ratios. An analogous expression
involving the lepton and slepton masses has been named $y_i^{ab}$. The
function coming from integrating over loop momentum is
$f(x,y)=[\log(y)/(y-1) - \log(x)/(x-1)]$. The $j$-sums run over all
squarks in $F_{u\tilde u}$, $F_{d\tilde d}$, and $F_{q\tilde q}$, and
over all sleptons in $F_{l\tilde l}$. The $i$-sums count all quarks in
$F_{q\tilde q}$, up-type quarks only in $F_{u\tilde u}$, down-type
quarks in $F_{d\tilde d}$, and all leptons in $F_{l\tilde l}$. The
factor 3 in $F_{u\tilde u}$, $F_{d\tilde d}$, and $F_{q\tilde q}$
accounts for summation over quarks' colors. It is absent in the case of
leptons.

We do not discuss the ${\cal M}^i$ contributions separately. The reason
is that for different cases the couplings $C_2$ and $C_3$ enter with
opposite signs causing cancellations between such terms. Since none of
the ${\cal M}^i$ can show up without the others, only the full sum Eq.
(\ref{eq:M-full}) gives a~meaningful picture.


\section{Phenomenological Majorana neutrino mass matrix}

The neutrino mass matrix can be constructed from the
Pontecorvo--Maki--Nakagawa--Sakata mixing matrix $U_{PMNS}$ under
certain assumptions. The matrix $U_{PMNS}$ is usually parameterized by
three angles and three (in the case of Majorana neutrinos) phases as
follows:
\begin{widetext}
\begin{equation}
  U_{PMNS} = 
  \left (
    \begin{array}{ccc}
      c_{12} c_{13} & s_{12} c_{13} & s_{13} e^{-i \delta} \\
      -s_{12} c_{23} - c_{12} s_{23} s_{13}e^{i \delta} & c_{12} c_{23} - s_{12}
      s_{23} s_{13}e^{i \delta} & s_{23} c_{13} \\
      s_{12} s_{23} - c_{12} c_{23} s_{13}e^{i \delta} & -c_{12} s_{23} - s_{12}
      c_{23} s_{13}e^{i \delta} & c_{23} c_{13}
    \end{array}
  \right ) 
  \left (
    \begin{array}{ccc}
      1 & 0 & 0 \\
      0 & e^{i \phi_{2}} & 0 \\
      0 & 0 & e^{i \phi_{3}} 
    \end{array}
  \right ),
  \label{U}
\end{equation}
where $s_{ij} \equiv \sin\theta_{ij}$, $c_{ij} \equiv \cos\theta_{ij}$.
Three mixing angles $\theta_{ij}$ ($i < j$) vary between 0 and $\pi/2$.
The $\delta$ is the CP violating Dirac phase and $\phi_2$, $\phi_3$ are
CP violating Majorana phases. Their values vary between $0$ and $2\pi$.
The explicit expression for the phenomenological mass matrix ${\cal
  M}^{ph}_{\alpha\beta}$ in terms of $m_i$, $\theta_{ij}$, $\delta$,
$\phi_2$, $\phi_3$ is given by \cite{mg-art9}:
\begin{eqnarray}
{\cal M}_{ee} &=&
~~c_{13}^{2}\,c_{12}^{2}\,{m_1}\,+\,
c_{13}^{2}\,s_{12}^{2}\,{m_2}\,{e^{-i2\phi_{_2}}}\,+\,
s_{13}^{2}\,{e^{2\,i\delta}}\,{m_3}\,{e^{-i2\phi_{_3}}},
\nonumber\\
\nonumber \\
{\cal M}_{e\mu} &=&
-\,c_{12}\,c_{13}\, 
\left( \,c_{23}\,s_{12}\,+\,c_{12}\,s_{23}\,s_{13}\,{e^{-i\delta}}\,
\right) {\it {m_1}}
\nonumber\\
&&+\,c_{13}\,s_{12}
\left( \,c_{23}\,c_{12}\,-\,s_{23}\,s_{12}\,s_{13}\,{e^{-i\delta}}
\right)\,{\it {m_2}} \,{e^{-i 2\phi_{_2}}} 
\,+\,c_{13}\,s_{23}\,s_{13}\,{e^{i\delta}}\,{m_3}\,{e^{-i2\phi_{_3}}},
\nonumber\\
\nonumber\\
{\cal M}_{e\tau} &=&
-\,c_{12}\,c_{13}\,
\left(-\,s_{23}\,s_{12}\,+\,c_{23}\,c_{12}\,s_{13}\,{e^{-i\delta}}
\right)\,{m_1}
\nonumber\\
&&-\,c_{13}\,s_{12}\,
\left(\,c_{12}\,s_{23}\,+\,c_{23}\,s_{12}\,s_{13}\,{e^{-i\delta}}\,
\right)
{m_2}{e^{-i2\phi_{_2}}}\,+\,
c_{23}\,c_{13}\,s_{13}\,{e^{i\delta}}\,{m_3}\,{e^{-i2\phi_{_3}}},
\nonumber\\
\nonumber\\
{\cal M}_{\mu\mu} &=&
~~\left(
c_{23}^2\,s_{12}^2\,+\,
2\,c_{23}\,c_{12}\,s_{23}\,s_{12}\,s_{13}\,{e^{-i\delta}}\,+
c_{12}^2\,s_{23}^2\,s_{13}^2\,{e^{-2\,\delta}}
\right)\,m_1\,
\nonumber\\
&&+\left(
c_{23}^2\,c_{12}^2\,-\,
2\,c_{23}\,c_{12}\,s_{23}\,s_{12}\,s_{13}\,{e^{-i\delta}}\,+\,
s_{23}^2\,s_{12}^2\,s_{13}^2\,{e^{-2\,\delta}}
\right) 
\,{m_2}\,{e^{-i2\phi_{_2}}} 
\,+\,
c_{13}^2\,
s_{23}^2\,
{m_3}\,{e^{-i2\phi_{_3}}},
\nonumber\\
\nonumber\\
{\cal M}_{\mu\tau} &=&
-\left( 
c_{23}\,s_{23}\,s_{12}^{2}\,
-\,c_{23}^{2}\,c_{12}\,s_{12}\,s_{13}\,{e^{-i\delta}}\,
+\,c_{12}\,s_{23}^{2}\,s_{12}\,s_{13}\,{e^{-i\delta}}\,
-\,c_{23}\,c_{12}^{2}\,s_{23}\,s_{13}^{2}\,{e^{-2\,i\delta}}\,
\right) {\it {m_1}}
\nonumber\\
&&- \left( 
c_{23}\,c_{12}^{2}\,s_{23}\,
+\,c_{23}^{2}\,c_{12}\,s_{12}\,s_{13}\,{e^{-i\delta}}\,
-\,c_{12}\,s_{23}^{2}\,s_{12}\,s_{13}\,{e^{-i\delta}}\,
-\,c_{23}\,s_{23}\,s_{12}^{2}\,s_{13}^{2}\,{e^{-2\,i\delta}}\,
\right) \,{m_2}\,{e^{-i2\phi_{_2}}}
\nonumber\\
&&\,+\,c_{23}\,c_{13}^{2}\,s_{23}\,{m_3}\,{e^{-i2\phi_{_3}}},
\nonumber\\
\nonumber\\
{\cal M}_{\tau\tau} &=&
~~\left( s_{23}^{2}\,s_{12}^{2}
\,-\,2\,c_{23}\,c_{12}\,s_{23}\,s_{12}\,s_{13}\,{e^{-i\delta}}
\,+\,c_{23}^{2}\,c_{12}^{2}\,s_{13}^{2}\,{e^{-2\,i\delta}}\right) {m_1}
\nonumber\\
&&+\left( 
c_{12}^{2}\,s_{23}^{2}\,+
\,2\,c_{23}\,c_{12}\,s_{23}\,s_{12}\,s_{13}\,{e^{-i\delta}}
+c_{23}^{2}\,s_{12}^{2}\,s_{13}^{2}\,{e^{-2\,i\delta}}
\right) 
\,{m_2}\,{e^{-i2\phi_{_2}}}\,
+\,c_{23}^{2}\,c_{13}^{2}\,{m_3}\,{e^{-i2\phi_{_3}}}.
\nonumber\\
\end{eqnarray}
\end{widetext}

In order to calculate numerical values of elements of this matrix one
needs some additional relations among the mass eigenstates
$m_{1,2,3}$. Experiments in which neutrino oscillations are investigated
allow to measure the absolute values of differences of neutrino masses
squared and the values of the mixing angles. The best-fit values of
these parameters read \cite{nu-osc,nu-mass}
\begin{eqnarray}
  |m_1^2 - m_2^2| &=& 7.1 \times 10^{-5} \eV^2, \nonumber\\
  |m_2^2 - m_3^2| &=& 2.1 \times 10^{-3} \eV^2, \nonumber\\
  \sin^2(\theta_{12}) &=& 0.2857,                          \\
  \sin^2(\theta_{23}) &=& 0.5,                    \nonumber\\
  \sin^2(\theta_{13}) &=& 0.                      \nonumber
\end{eqnarray}
The present experimental outcomes are in agreement with two
scenarios:
\begin{itemize}
\item the \emph{normal hierarchy} (NH) of masses imply the relation 
  $m_1 < m_2 < m_3$,
\item the \emph{inverted hierarchy} (IH) of masses imply the relation
  $m_3 < m_1 < m_2$.
\end{itemize}
Notice that in order to keep the same notation for the differences of
masses squared and the mixing angles, the neutrino mass eigenstates are
labeled differently in the NH and IH cases.

At this point we are left with four undetermined parameters, which are
the phases and the mass of the lightest neutrino. To obtain most
stringent limits on the new physics parameters the later is taken to be
zero. As far as the phases are concerned we consider two separate cases.
First we take all possible combinations of phases and for each entry of
the matrix we pick up its highest possible value. In this way we obtain
unphysical matrices, which give however some idea about the upper limits
on the non-standard parameters. The maximal matrices for the NH and IH
scenarios read as follows \cite{mg-art9}:

\begin{equation}
|\cal{M}|_{\rm max}^{\rm (NH)} \le
\pmatrix{ 
  0.00452 & 0.00989 & 0.00989 \cr
  0.00989 & 0.02540 & 0.02540 \cr
  0.00989 & 0.02540 & 0.02540 }    \eV,
\end{equation}

\begin{equation}
|\cal{M}|_{\rm max}^{\rm (IH)} \le
\pmatrix{ 
  0.0452 & 0.0312 & 0.0312 \cr
  0.0312 & 0.0240 & 0.0239 \cr
  0.0312 & 0.0239 & 0.0240 }       \eV.
\end{equation}

The more conservative approach assumes that the $CP$ symmetry is
preserved which can be achieved by neglecting the phases present in the
$U_{PMNS}$ matrix. In such a~case the NH and IH matrices take the
following forms:

\begin{equation}
|\cal{M}|^{\rm (NH)} = 
\pmatrix{
  0.00240 & 0.00269 & 0.00269 \cr
  0.00269 & 0.02553 & 0.01951 \cr
  0.00269 & 0.01951 & 0.02553 }   \eV,
\end{equation}

\begin{equation}
|\cal{M}|^{\rm (IH)} = 
\pmatrix{ 
  0.045267 & 0.000249 & 0.000249 \cr
  0.000249 & 0.022801 & 0.022801 \cr
  0.000249 & 0.022801 & 0.022801 }   \eV.
\end{equation}

Yet another possibility is to construct ${\cal M}$ using constraints
from non-observability of the neutrinoless double beta decay
($0\nu2\beta$). The study of the $0\nu2\beta$ decay \cite{0n2b} is one
of the most sensitive ways known to probe the absolute values of
neutrino masses and the type of the spectrum. The most stringent lower
bound on the half-life of $0\nu2\beta$ decay were obtained in the
Heidelberg-Moscow $^{76}$Ge experiment \cite{H-M}
($T^{0\nu-exp}_{1/2}\ge 1.9\times 10^{25}$ yr). By assuming the nuclear
matrix element of Ref.~\cite{FedorVogel} we end up with
$|m_{\beta\beta}| =U^{2}_{e1}\,m_{1} + U^{2}_{e2}\,m_{2}
+U^{2}_{e3}\,m_{3} \le 0.55 \eV$, where $U$ is the neutrino mixing
matrix Eq.~(\ref{U}). The element $|m_{\beta\beta}|$ coincides with the
$ee$ element of the neutrino mass matrix in the flavor basis and fixing
it allows to construct the full maximal matrix, which reads:
\begin{eqnarray}
  |{\cal M}|_{\rm max}^{\rm (HM)} \le 
  \pmatrix{
    0.55 & 1.29 & 1.29 \cr
    1.29 & 1.35 & 1.04 \cr
    1.29 & 1.04 & 1.35 }  \eV.
\end{eqnarray}
In the next section we present the results for each of these five
cases. 

\begin{table*}
  \centering
  \caption{\label{tab:2} Some results for the SUSY scenario $\tan\beta=10$,
    $A_0=200$, $m_0=200\GeV$, $m_{1/2}=500\GeV$. } 
  \begin{tabular}{cccccl}
    \hline\hline
    $\kappa_1^{GUT}$ & $\kappa_2^{GUT}$ & $\kappa_3^{GUT}$ & 
    Resulting mass matrix & Compare with & Remarks \\
    \multicolumn{3}{c}{[MeV]} & [eV] & \phantom{xxxxxxxxxxxxxxxxx} & \\
    \hline
    $9.50$ & $14.80$ & $14.80$ &
    $\pmatrix{  
      0.553927 & 0.864372 & 0.861090  \cr
      0.865038 & 1.349844 & 1.344714  \cr
      0.859783 & 1.341638 & 1.336555  }$  & ${\cal M}^{\rm (HM)}_{\rm max}$
    & $\mu\tau$ elements to big \\
    $9.46$ & $13.02$ & $13.02$ &
    $\pmatrix{  
      0.548762 & 0.757606 & 0.753587  \cr
      0.757853 & 1.046272 & 1.040721  \cr
      0.752398 & 1.038740 & 1.033237  }$  & ${\cal M}^{\rm (HM)}_{\rm max}$
    & \\
    $0.85$ & $2.03$ & $2.03$ &
    $\pmatrix{  
      0.004520 & 0.010728 & 0.010698 \cr
      0.010734 & 0.025474 & 0.025404 \cr
      0.010686 & 0.025361 & 0.025292 }$  & ${\cal M}^{\rm (NH)}_{\rm max}$
    & $e\mu$ and $e\tau$ elements to big \\
    $2.72$ & $1.98$ & $1.98$ &
    $\pmatrix{  
      0.045316 & 0.032954 & 0.032976 \cr
      0.032945 & 0.023958 & 0.023974 \cr
      0.032951 & 0.023963 & 0.023978 }$  & ${\cal M}^{\rm (IH)}_{\rm max}$
    & $e\mu$ and $e\tau$ elements to big \\
    $0.62$ & $2.03$ & $2.03$ &
    $\pmatrix{  
      0.002402 & 0.007824 & 0.007802  \cr
      0.007821 & 0.025474 & 0.025404  \cr
      0.007787 & 0.025361 & 0.025292  }$  & ${\cal M}^{\rm (NH)}$
    & $e\mu$, $e\tau$ and $\mu\tau$ elements to big \\
    $0.62$ & $0.70$ & $0.70$ &
    $\pmatrix{  
      0.002402 & 0.002691 & 0.002687 \cr  
      0.002688 & 0.003011 & 0.003007 \cr  
      0.002684 & 0.003007 & 0.003003  }$  & ${\cal M}^{\rm (NH)}$
    \\
    $2.72$ & $1.92$ & $1.93$ &
    $\pmatrix{  
      0.045316 & 0.032053 & 0.032118  \cr
      0.032065 & 0.022680 & 0.022726  \cr
      0.032084 & 0.022694 & 0.022740  }$  & ${\cal M}^{\rm (IH)}$
    & $e\mu$ and $e\tau$ elements to big \\
    $0.27$ & $0.19$ & $0.19$ &
    $\pmatrix{  
      0.000453 & 0.000320 & 0.000321  \cr
      0.000320 & 0.000226 & 0.000227  \cr
      0.000320 & 0.000226 & 0.000227  }$  & ${\cal M}^{\rm (IH)}$
    \\ \hline\hline
  \end{tabular}
\end{table*}


\section{Constraining $\kappa$ couplings from the neutrino mass matrix}

\begin{figure}
  \centering
  \includegraphics[width=0.45\textwidth]{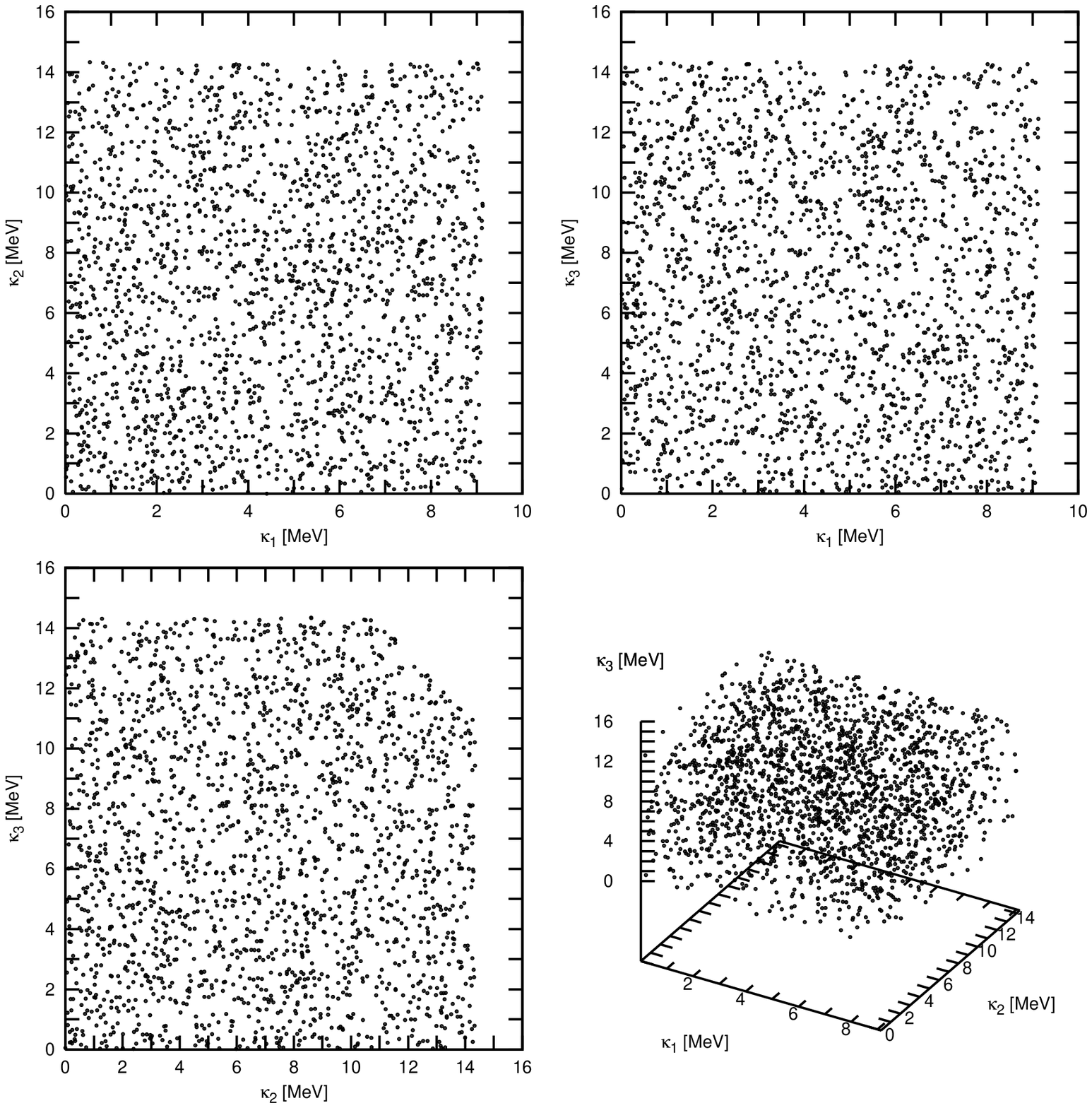}
  \caption{\label{fig:HM} Allowed parameter space in the maximal HM
    case.}
\end{figure}

\begin{figure}
  \centering
  \includegraphics[width=0.45\textwidth]{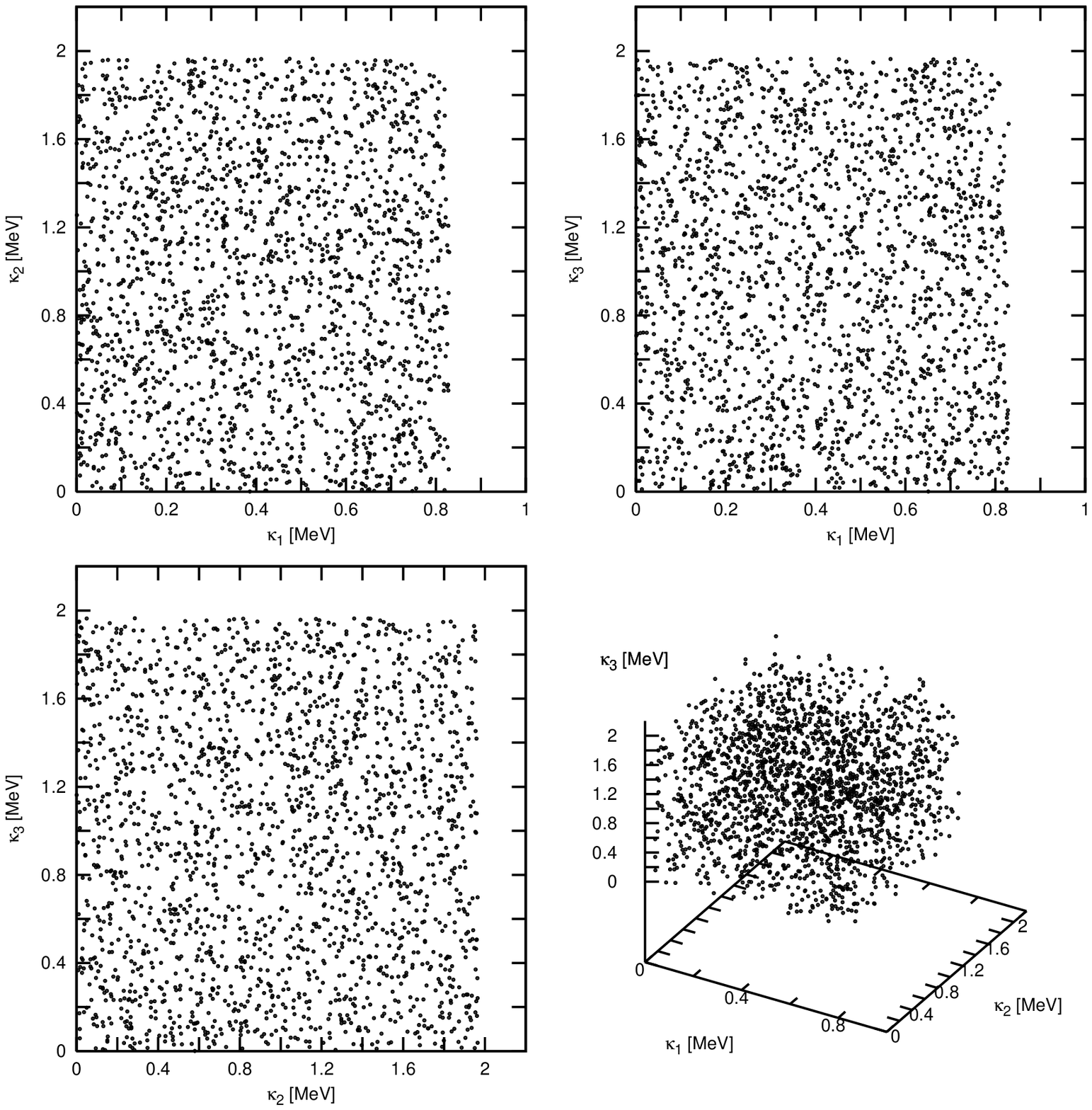}
  \caption{\label{fig:NHmax} Allowed parameter space in the maximal NH
    case.}
\end{figure}

\begin{figure}
  \centering
  \includegraphics[width=0.45\textwidth]{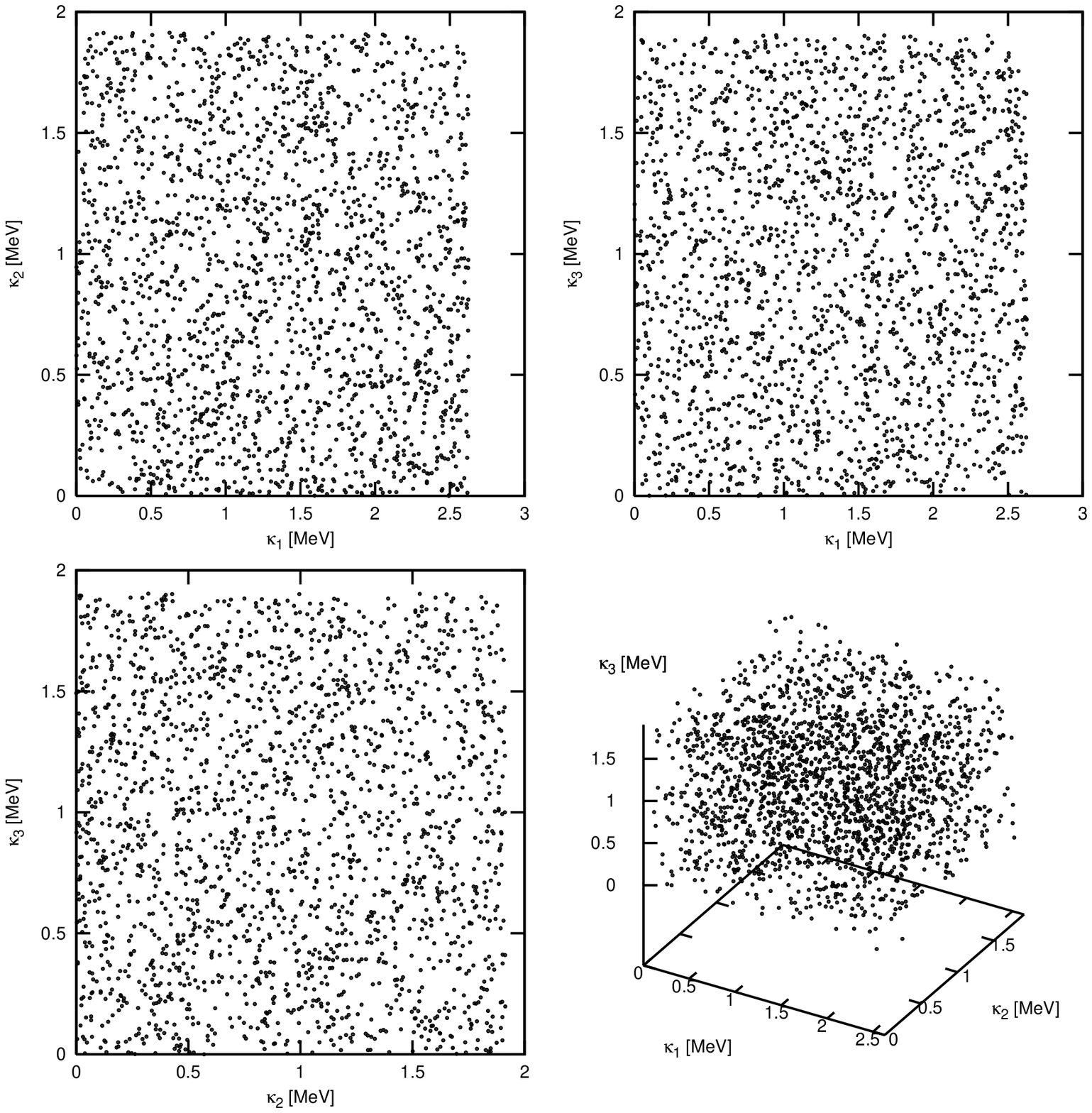}
  \caption{\label{fig:IHmax} Allowed parameter space in the maximal IH
    case.}
\end{figure}

\begin{figure}
  \centering
  \includegraphics[width=0.45\textwidth]{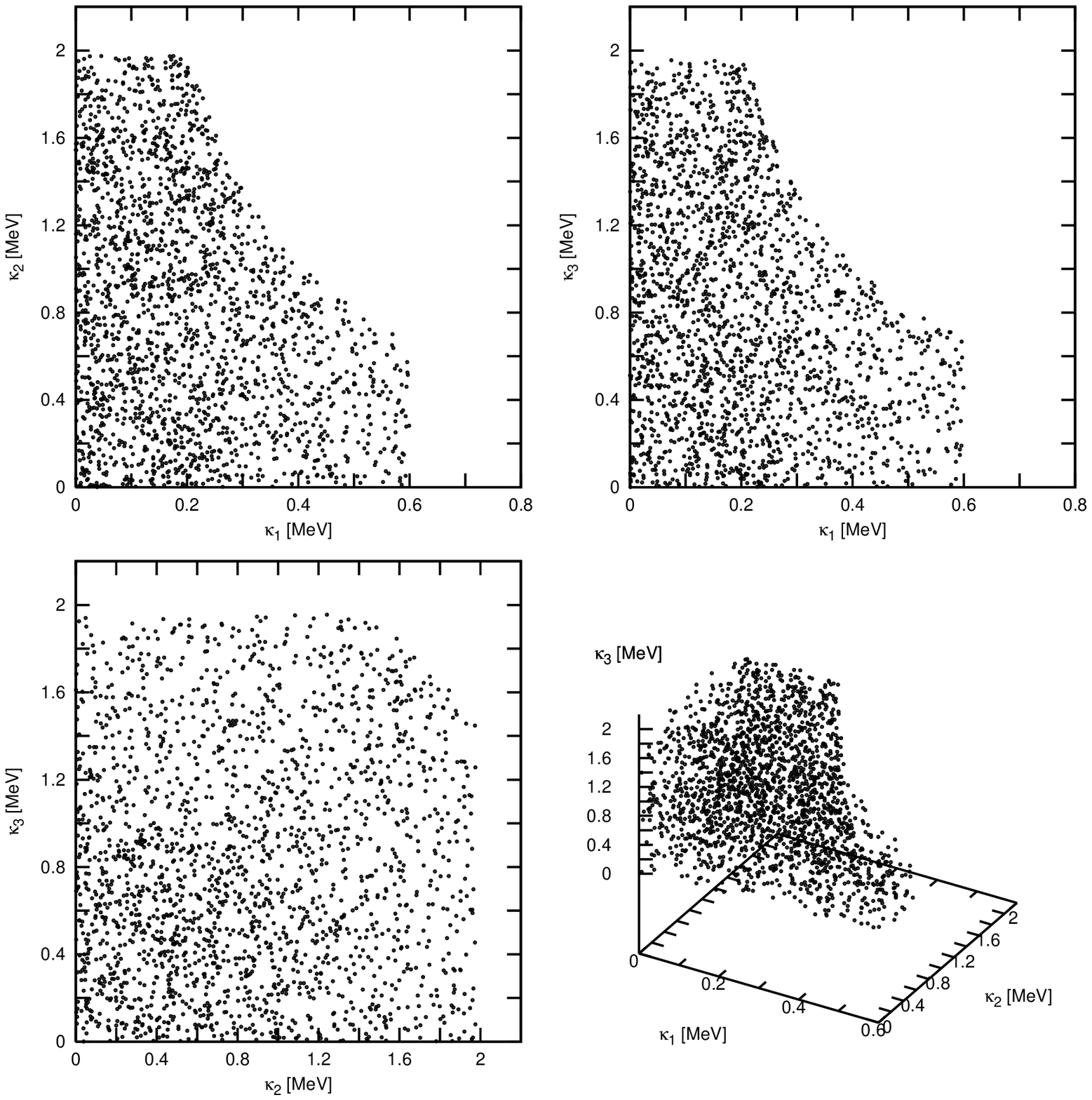}
  \caption{\label{fig:NHCP} Allowed parameter space in the NH case with
    conserved CP symmetry.}
\end{figure}

\begin{figure}
  \centering
  \includegraphics[width=0.45\textwidth]{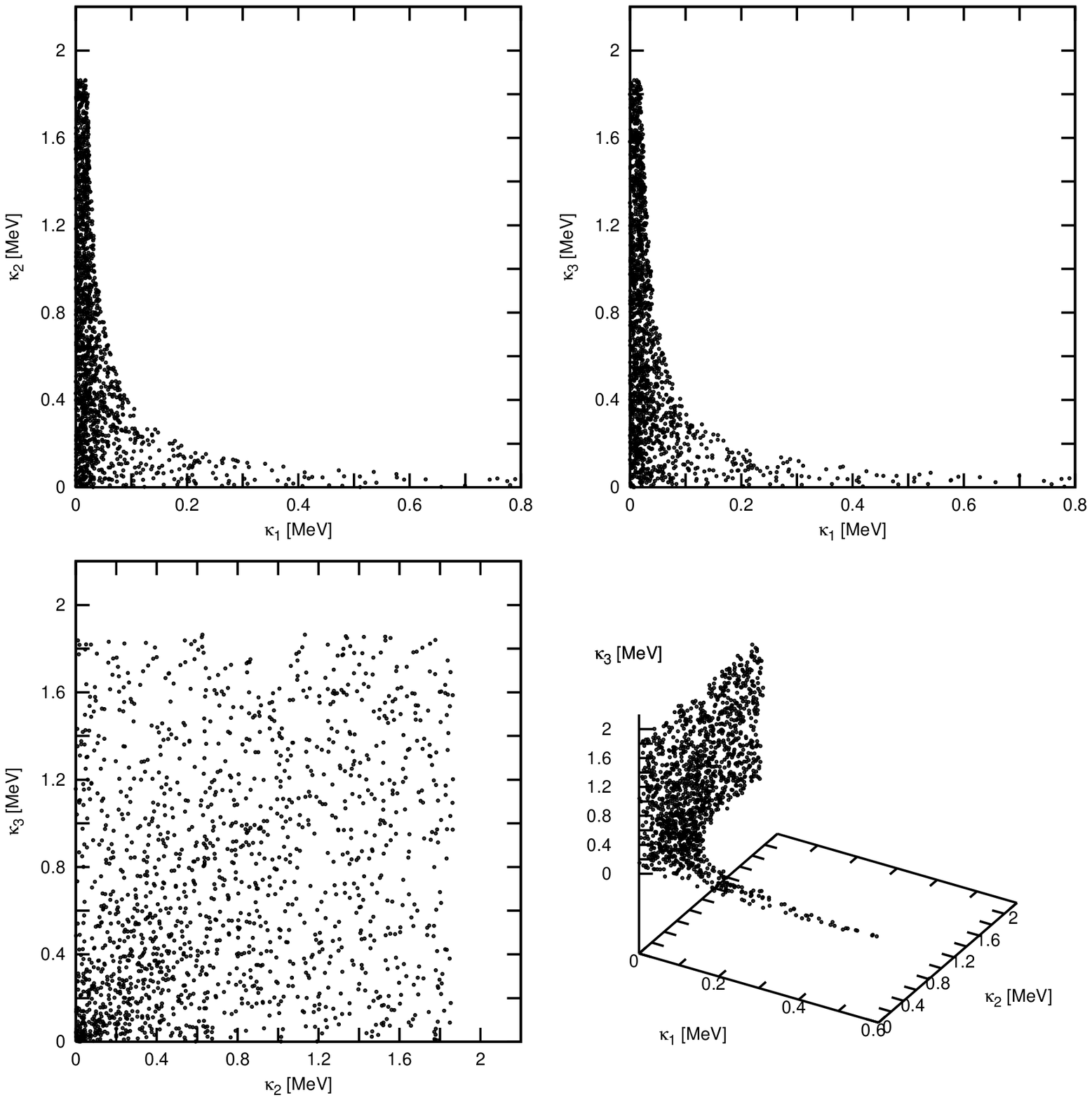}
  \caption{\label{fig:IHCP} Allowed parameter space in the IH case with
    conserved CP symmetry.}
\end{figure}

Our aim is to find constraints on the $\kappa_i$ coupling constants
coming from the neutrino mass matrices. As an example of the unification
conditions we take the following input:
\begin{equation}
  A_0 = 200, \quad m_0 = 200 \GeV, \quad m_{1/2} = 500 \GeV,
\end{equation}
and additionally
\begin{equation}
  \tan\beta = 10, \qquad {\rm sgn}(\mu) = 1.
\end{equation}
We do not expect great differences in the results if the GUT conditions
were changed. The only exception may be the $\tan\beta$ parameter
(defined at $m_z$ scale) which dominates the running of the
$\kappa$'s. By looking on Fig.~\ref{fig:kappa} only very low values of
this parameter will influence the results significantly.

We proceed in two steps. Firstly we find such values of $\kappa_i^{GUT}$
which will reproduce the diagonal elements of the mass matrices. This
can be achieved with good accuracy, but it turns out that some of the
elements (off-diagonal, see remarks in Tab.~\ref{tab:2}) of the
resulting matrix exceed the allowed values. It means that the $\kappa$'s
will not take their maximal values simultaneously.

Secondly we go down with the $\kappa_i^{GUT}$ to lower the off-diagonal
elements to the acceptable level. This, however, can be done in many
different ways. Some explicit examples are listed in Tab.~\ref{tab:2}
but to find the full allowed parameter space we have prepared scatter
plots which are presented on Figs.~\ref{fig:HM}--\ref{fig:IHCP}. Each of
the plots consists of roughly 2000 points chosen randomly from the
intervals between zero and 1.1 times the assessed upper limit for given
$\kappa_i^{GUT}$.

The boundaries of the allowed parameter space for $\kappa_i$ in the case
of unphysical neutrino mass matrices $|{\cal M}|_{\rm max}^{\rm (HM, NH,
  IH)}$ are nearly box-shaped, except for a~small region of excluded
values in the upper right-hand corner of some of the plots. This
behavior is expected, as the mass matrices to be reproduced contain on
each entry the maximal allowed value for it.

Much more interesting shapes are obtained for the CP conserved cases
(Figs.~\ref{fig:NHCP} and \ref{fig:IHCP}). The projections onto the
$(\kappa_1,\kappa_2)$ and $(\kappa_1,\kappa_3)$ planes are nearly
identical and contain non-linear boundary parts. The most constrained is
the inverted hierarchy case with conserved CP (Fig.~\ref{fig:IHCP}) due
to two orders of magnitude differences between the diagonal $\mu\mu$ and
$\tau\tau$ and off-diagonal $e\mu$ and $e\tau$ elements.


\section{The transition magnetic moment}
%
\begin{figure}[b]
  \includegraphics[width=0.35\textwidth]{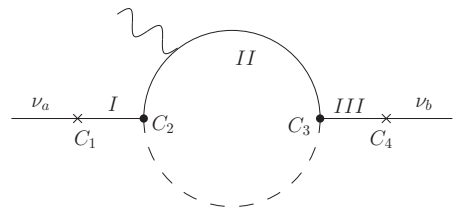}
  \caption{\label{fig:diagram2} Feynman diagram with
    neutrino-neutralino mixing on the external lines, leading to the
    Majorana neutrino transition magnetic moment.}
\end{figure}
%
The RpV loop diagrams provide not only an elegant mechanism of
generating Majorana neutrino mass terms, but also, after a~minor
modification, may be the source of the transition magnetic moment
$\mu_{ab}$. This quantity represents roughly the strength of the
electromagnetic interaction of the neutrino. Since the latter is
electrically neutral, the interaction must take place between an
external photon and a~charged particle from inside the virtual RpV
loop. In practice, only the photon--fermion interactions are taken into
account, since the photon--boson (squark or slepton) interaction would
be strongly suppressed by the big mass of the SUSY particle. The
relevant Feynman diagram is presented on Fig.~\ref{fig:diagram2}.

The contribution to the Majorana neutrino magnetic moment from the
discussed diagrams is given by (in Bohr magnetons $\mu_B$)
\begin{eqnarray}
  \mu_{ab} &=& (1-\delta_{ab}) \frac{m_{e^1}}{4\pi^2}
  \ \left(C_{1a}
    \frac{C_2 C_3}{m_I m_{III}}
    C_{4b} \right ) \nonumber \\ 
  &\times& \sum_{i,j}
  \left [
    3\frac{w_{ij}^{(q)}}{m_{q^i}} Q_{q^i} +
    \frac{w_{ij}^{(\ell)}}{m_{\ell^i}} Q_{\ell^i}
  \right ] \mu_B.
\label{eq:mu}
\end{eqnarray}
Here we have denoted the electric charge of a~particle (in units of $e$)
by $Q$. The dimensionless loop functions $w$ take the forms
\begin{equation}
  w_{ij}^{(q)} = \frac{\sin2\theta^j}{2} g(x_2^{ij},x_1^{ij}), \quad
  w_{ij}^{(\ell)} = \frac{\sin2\phi^j}{2} g(y_2^{ij},y_1^{ij}),
\label{eq:w}
\end{equation}
where $\theta$, $\phi$, $x_{1,2}$, and $y_{1,2}$ are the same as in
Eqs.~(\ref{eq:Fu})--(\ref{eq:Fl}), and $g(x,y)=(x\log(x)-x+1)(1-x)^{-2}
- (x\to y)$. The sum over $i$ and $j$ in Eq.~(\ref{eq:mu}) accounts for
all the possible quark-squark and lepton-slepton configurations for
given neutralinos. The factor 3 in front of $w^{(q)}$ counts the three
quark colors.

The results for the already discussed GUT parameters are presented
in Tab.~\ref{tab:3}. The last column contains for comparison upper
bounds for the magnetic moment in the case when only trilinear
interactions are taken into account. One sees that they are at least one
order of magnitude stronger than the discussed bilinear contributions.
%
\begin{table}
  \caption{\label{tab:3}Contribution to the Majorana neutrino transition
    magnetic moments coming from the bilinear neutrino-neutralino mixing,
    for the GUT scenario: $A_0=200$, $m_0=200\GeV$, $m_{1/2}=500\GeV$,
    $\tan\beta=10$.}
\begin{tabular}[c]{lcccc}
  \hline\hline
  &  $\mu_{e\mu}$      & $\mu_{e\tau}$      & $\mu_{\mu\tau}$ & trilinear only\\
  \hline
  IH-CP  &$7.0\times 10^{-22}$&$7.0\times 10^{-22}$&$6.0\times 10^{-20}$&$\le 10^{-19}$\\
  IH-max &$8.8\times 10^{-20}$&$8.5\times 10^{-20}$&$6.5\times 10^{-20}$&$\le 10^{-17}$\\
  \\
  NH-CP  &$7.6\times 10^{-21}$&$7.6\times 10^{-21}$&$5.5\times 10^{-20}$&$\le 10^{-18}$\\
  NH-max &$2.8\times 10^{-20}$&$2.8\times 10^{-20}$&$7.0\times 10^{-20}$&$\le 10^{-17}$\\
  \\
  HM-max &$2.4\times 10^{-18}$&$2.3\times 10^{-18}$&$2.9\times 10^{-18}$&$\le 10^{-15}$\\
  \hline\hline
\end{tabular}
\end{table}
%
\begin{table}
  \caption{\label{tab:4}Like in Tab.~\ref{tab:3} but with $A_0=100$,
    $m_0=150\ \GeV$, $m_{1/2}=150\ \GeV$, $\tan\beta=19$.}
\begin{tabular}[c]{lcccc}
  \hline\hline
  &  $\mu_{e\mu}$        & $\mu_{e\tau}$      & $\mu_{\mu\tau}$ & trilinear only\\
  \hline
  IH-CP  &  $3.0\times 10^{-21}$&$2.9\times 10^{-21}$&$2.5\times 10^{-19}$&$\le 10^{-18}$\\
  IH-max &  $3.7\times 10^{-19}$&$3.6\times 10^{-19}$&$2.7\times 10^{-19}$&$\le 10^{-18}$\\
  \\
  NH-CP  &  $3.2\times 10^{-20}$&$3.1\times 10^{-20}$&$2.2\times 10^{-19}$&$\le 10^{-18}$\\
  NH-max &  $1.2\times 10^{-19}$&$1.1\times 10^{-19}$&$2.9\times 10^{-19}$&$\le 10^{-18}$\\
  \\
  HM-max &  $1.0\times 10^{-17}$&$9.8\times 10^{-18}$&$1.2\times 10^{-17}$&$\le 10^{-16}$\\
  \hline\hline
\end{tabular}
\end{table}
%
\begin{table}
  \caption{\label{tab:5}Like in Tab.~\ref{tab:3} but with $A_0=500$,
    $m_0=1000\ \GeV$, $m_{1/2}=1000\ \GeV$, $\tan\beta=19$.}
\begin{tabular}[c]{lcccc}
  \hline\hline
  &  $\mu_{e\mu}$        & $\mu_{e\tau}$      & $\mu_{\mu\tau}$ & trilinear only \\
  \hline
  IH-CP  &  $3.7\times 10^{-22}$&$3.7\times 10^{-22}$&$3.3\times 10^{-20}$&$\le 10^{-20}$\\
  IH-max &  $4.6\times 10^{-20}$&$4.6\times 10^{-20}$&$3.5\times 10^{-20}$&$\le 10^{-20}$\\
  \\
  NH-CP  &  $4.0\times 10^{-21}$&$4.0\times 10^{-21}$&$2.9\times 10^{-21}$&$\le 10^{-20}$\\
  NH-max &  $1.4\times 10^{-20}$&$1.5\times 10^{-20}$&$3.7\times 10^{-20}$&$\le 10^{-20}$\\
  \\
  HM-max &  $1.3\times 10^{-18}$&$1.2\times 10^{-18}$&$1.5\times 10^{-18}$&$\le 10^{-18}$\\
  \hline\hline
\end{tabular}
\end{table}
%
In Tabs.~\ref{tab:4} and \ref{tab:5} we show the results of similar
calculations for two other set of parameters. In Tab.~\ref{tab:4} the
unification parameters are `low', while in Tab.~\ref{tab:5} their values
are `higher'. The last column is given as previously for comparison.
Also here the conclusion is clear, that the discussed contribution to
the main process is at best of the same order of magnitude, in most
cases being at least an order of magnitude weaker. The reason for such
a~situation is due to the high masses of the neutralinos, which enter
the formula Eq.~(\ref{eq:mu}) in the denominator. It was possible that
they will be compensated by the unknown coupling constants proportional
to the sneutrino vacuum expectation values $\omega$, especially that
some of them are multiplied by negative numbers, cf. Tab.~\ref{tab:2}.
Our explicit calculation showed that it is not the case. Also the
observed differences between the values of $\mu_{ab}$, reaching not more
than one order of magnitude, are mainly due to the changed value of the
parameter $\tan\beta$, and only partially due to different values of the
remaining parameters.


\section{Summary}

The $R$-parity violating MSSM has many free parameters which lower its
predictive power. On the other hand this fact makes the model very
flexible. In this paper we have presented a method of constraining the
bilinear RpV couplings $\kappa$.

We have calculated the contributions to the neutrino mass matrix coming
from the neutrino--neutralino mixing in processes in which the effective
vertex is expanded into a virtual quark--squark or lepton--slepton
loop. These contributions have been compared with the phenomenological
mass matrices derived using the best-fit experimental values of the
neutrino mixing angles and differences of masses squared. We discuss
four cases in which normal and inverted hierarchy is explored both with
conserved CP symmetry and with maximal values of each matrix element. We
also present the fifth case in which the neutrino mass matrix is
calculated from the data published by the Heidelberg--Moscow
neutrinoless double beta decay experiment.

In general we have found that setting the $\kappa$ couplings at the
unification scale to values of the order of $\lesssim {\cal O}(1\MeV)$
renders the mass contributions correctly below the experimental upper
bound. Another observation is that the bilinear RpV mechanism alone is
not sufficient to reproduce the whole mass matrix. This is, however,
acceptable because in the general RpV loop mechanism one has to sum up
the contributions from the tree--level \cite{Haug},
\begin{eqnarray}
  && {\cal M}^{tree}_{ii'} = \Lambda_i \Lambda_{i'} \ g_2^2 \cr
  &&\times  \frac{M_1 + M_2 \tan^2\theta_W}
  {4(\mu m_W^2 (M_1 + M_2 \tan^2\theta_W)\sin2\beta - M_1 M_2\mu^2)},
  \cr &
\label{Mtree}
\end{eqnarray}
where $\Lambda_i = \mu \omega_i - v_d \kappa_i$ are the so-called
alignment parameters, as well as contributions coming from the 1--loop
diagrams (see Fig.~\ref{fig:loop}(a)), which are proportional to the
totally unconstrained trilinear couplings $\lambda$ and $\lambda'$.
These parameters may be easily fine-tuned to reproduce the full mass
matrix and we shift this discussion to an upcoming paper.

The knowledge of the bounds on the $\kappa$ coupling constants allows
one to discuss many exotic processes, like the neutrino decay and the
interaction of neutrino with a~photon, to mention only a~few. The former
may occur as a~two-step process, first through bilinear mixing with
neutralinos, and then the decay of the actual neutralino. The later has
been presented in the previous section showing by explicit calculation
that this contribution does not exceed the main 1--loop mechanism.

In our calculations we have fixed the GUT unification parameters. Due to
technical difficulties in performing a~full skan over the allowed
parameter space we have picked only three representatives for which the
calculations were performed. We expect that the results will not change
qualitatively with the changes of the input parameters, which is of
course an assumption that may be worth checking.


\section*{Acknowledgments}

The first author (MG) is partially supported by the Polish State
Committee for Scientific Research. He would like also to express his
gratitude to prof.~A.~F\"a{\ss}ler for his warm hospitality in
T\"ubingen during the Summer 2006.



\end{document}